%% file: sample-sigconf.tex
\documentclass[sigconf,10pt]{acmart}
\renewcommand\footnotetextcopyrightpermission[1]{}
\usepackage{subcaption}
\usepackage{tabularx}
\usepackage{hyperref}
\usepackage[inline]{enumitem}
\setlist[itemize]{leftmargin=*}

\usepackage{booktabs} 

\begin{document}
\title{Playing with Matches: Vehicular Mobility through Analysis of Trip Similarity and Matching}

\author{Roozbeh Ketabi}
\affiliation{%
  \institution{University of Florida}
  \city{Gainesville}
  \state{Florida}
}
\email{roozbeh@ufl.edu}

\author{Babak Alipour}
\affiliation{%
  \institution{University of Florida}
  \city{Gainesville}
  \state{Florida}
}
\email{babak.ap@ufl.edu}

\author{Ahmed Helmy}
\affiliation{%
  \institution{University of Florida}
  \city{Gainesville}
  \state{Florida}
}
\email{helmy@ufl.edu}


\begin{abstract}
Understanding city-scale vehicular mobility and trip patterns is essential to addressing many problems, from transportation and pollution to public safety, among others.
Using spatio-temporal analysis of vehicular mobility, promising solutions can be
proposed to alleviate these major challenges, utilizing shared mobility and
crowd-sourcing. The rise of transportation networks (e.g. Uber, Lyft),
is a mere beginning to shared mobility. Meanwhile, the ubiquity of
network-enabled devices and advances in data collection enables a
data-driven study of trips.
In this paper, we address problems of trip representation and matching.
Particularly, we study a real-world dataset of trips (from Cologne, Germany),
from spatial and temporal perspectives.
Comparison of trajectories is desired for applications relying on spatio-temporal phenomena. For that purpose, we
present a novel combined spatio-temporal similarity score, based on the weighted
geometric mean (WGM) and conduct experiments on its applicability and strengths.
First, we use the score to find clusters of trips that were spatially and/or temporally
separable using spectral clustering. The score is then used in a real-time matching of trips for Catch-a-Ride (CaR) and CarPooling (CP) scenarios. CaR and CP achieve $\approx40\%$ and $\approx25\%$ decrease in traveled distances respectively, at the cost of moving to pick-up and from drop-off locations (i.e. drivers going on average $<700m$ out of their way on pick-up and drop-off for CP).
Additionally, a comparison with the metrics available in the literature is
presented on CaR scenario. We find that main advantages of WGM include the flexibility to favor time or space components, and linearity of runtime complexity.
Finally, we formulate an optimal free float Car-Sharing scenario (e.g. scheduling a system of automated vehicles or taxis) resulting in an average of $\approx3.88$ trips serviced by a car in one hour.
\end{abstract}

%
%



\maketitle

\input{PlayMatches_body}

\bibliographystyle{ACM-Reference-Format}
\bibliography{PlayMatches_bibliography}

\end{document}

%% file: PlayMatches_body.tex
\section{Introduction}
Many problems plague big cities, including transportation, pollution, security, and public safety. With the proliferation of mobile devices and availability of more datasets, there is great potential to revisit these problems and introduce more efficient solutions. As a start, in this study, we shall focus on shared transportation as a solution to congestion problems in big cities.
Recent advances in mobile technology enabled the creation and rise of transportation network companies, like Uber and Lyft. Although shared transportation is not new, these companies were able to realize them at scale. Unfortunately, these companies, in order to maintain their competitive edge, do not disclose much of their data or analysis hence we investigate the matter with open datasets.
The approach to the study of human mobility depends on the spatial scale of the target varying from individual mobility to higher levels (buildings, city blocks, etc.). As vehicular transportation is the quintessence of urban mobility we focus on the study of trips happening in a city.
A common representation of trips is through their origins and destination. In this paper, we use origin-destination (OD) representation to study the trips and investigate various scenarios of shared transportation. A comparison of similarity/distance scores is then presented using a more detailed representation of trips based on a sample of 50 waypoints (spatio-temporal triples of x,y,t) for each trip (instead of OD - endpoints only).
Although similarity of trajectories has long been a subject of interest, the concept of time and its effects are not well-studied with respect to the trajectories. We approach this problem using spatio-temporal similarity of waypoints of the trips.
Public transportation is a big part of large cities around the world, but fine-grained shared mobility has received less attention. Recently, it has been seen for its potentials in the reducing travel times and traffic leading to increased productivity, less fuel consumption, and less pollution. In this paper, we use our proposed similarity measure to formulate and investigate three scenarios of shared transportation.

This study provides the following main contributions:
\begin{enumerate*}
\item Analysis of trips data from a spatio-temporal perspective,
\item Proposing a measure of spatio-temporal similarity of trajectories (trips) with flexibility across spatial or temporal elements. A comparison is conducted of various measures available in the literature,
\item Study of various shared transportation scenarios of Catch-a-Ride and Car Pooling. In addition, a novel graph formulation of a free float Car-Sharing scenario is presented. 
\end{enumerate*}

The study finds that trip distance and duration follow lognormal and gamma distributions (with gamma providing a better fit), and demonstrates the utility of the score and spectral clustering in finding spatio-temporal clusters. Similarity measures available in the literature have a quadratic computation complexity (in the number of waypoints), while our score has linear complexity with the flexibility to prefer time over space or vise versa. Shared transportation achieves roughly 25\% decrease in distances traveled by cars. The Car-Sharing scenario is optimal in the number of cars (minimum) and also the consecutive pick-ups in terms of combined temporal and spatial distances with the algorithm being bounded by the complexity of the matching in bipartite graphs ($O(n^3)$ in the number of nodes - in our case number of trips).

The rest of the paper is structured as follows: Related work in Sec. \ref{sec:relatedwork} followed by exploratory data analysis in Sec. \ref{sec:analysis}, and definition of similarity score and its potential in finding spatio-temporal clusters in Sec. \ref{sec:similarity} and \ref{sec:clustering}. The score is used for dynamic matching of trips for Catch-a-Ride and Car Pooling scenarios of shared transportation in Sec. \ref{sec:matching}. Sec. \ref{sec:compare} presents a comparison of various measures of similarity. Formulation of the Car Sharing problem and the algorithm is presented in Sec. \ref{sec:carsharing}. Lastly, Sec. \ref{sec:conclusion} concludes the study.

\vspace{-5pt}
\section{Related Work}
\label{sec:relatedwork}
We explore the literature from three perspectives: 
\begin{enumerate*}
\item The study of trips for their characteristics in time and space.
\item The similarity of trips using the trajectory of movements.
\item The concept of shared transportation and its variations.
\end{enumerate*}

Several studies use publicly available taxicab datasets to characterize trips.
\citet{Ferreira2013} analyzed a large dataset of taxicab trips (170M trips a year, for 2 years) of NYC for visualization of multiple use cases such as activity in different regions. 
Users trips have been extracted from geo-tagged images and studied for patterns of visitations \cite{arase2010mining}. A recent study focused on the travel patterns arising from Internet-based ride-sharing platforms using data from DiDi, China's biggest ride-hailing company \cite{Dong2018}. They found two types of individual behavior: commuters and taxi-like roaming.
In a study of spatial patterns based on the origin-destination representation of trips, \citet{Guo2012} proposed a distance measure based on the number of shared nearest neighbors and used it to find spatial clusters of the trips. We follow a similar approach but we define the similarity score and clustering based on temporal aspect in addition to space.
\citet{Veloso2011} studied a taxi dataset for Lisbon, Portugal. They visualized and explored various spatio-temporal features including distributions of trip durations, distances, pickups, and drop-offs.
Uber and Lyft trips in San Francisco were studied and visualized by SF County Transportation Authority with respect to location and time \cite{sfTNC}.

Trip trajectories have been studied for storage and performance optimizations, as well as traffic insights.
The similarity of trips bounded by traffic network was studied in \cite{Abraham2012}, using spatial and temporal points of interest and the distance of the trip to them. 
\citet{tiakas2006trajectory} defined spatial similarity between two trajectories as the average point-wise distance of points in a trajectory and temporal similarity as the sum of differences of consecutive points relative to the maximum of both.
\citet{Toohey2015} performed a comparative study of various similarity measures. They consider measures of longest common subsequence (LCSS), Frechet distance, dynamic time warping (DTW) and edit distance for trajectories of delivery drivers in the UK. These metrics are detailed in section \ref{sec:compare} for comparison.
In a similar study, \citet{Wang2013} investigated the effectiveness of various similarity scores on multiple transformations of trajectories, with parameters controlling the amplitude of change in the trips.
\citet{VanKreveld2007} studied the similarity of trajectories and subtrajectories considering time shifts, to find most similar trajectories. They propose various polynomial time algorithms; $O(n^4)$ for the most sophisticated case with time shifts and subtrajectories.
\citet{yan2017itdl} proposed a method of generating embeddings of Points of Interest (POI) with a spatial context that can be used to measure the similarity of place types. This has potential applications in the matching of rides and profiling trajectories.

Concerning rapidly expanding ride-hailing networks, \citet{Tian2013} presented a system of dynamic ride-sharing called Noah where three algorithms of matching are considered. 
Minimizing traveled distances or times along with maximizing number of matches are common non-commercial objectives of dynamic ride-sharing systems \cite{AGATZ2012295,DiFebbraro2013}.
\citet{Gidofalvi2008} studied the social barriers of shared mobility and proposed a system of ride sharing that incorporates social relationships in the matching criteria.
\citet{Masoud2017} studied another variant of ride-sharing, assuming a p2p (similar to carpooling) and flexible (multi-hop with transfers) scenario and provides an optimal solution. 
\citet{Ma2013} propose a framework of large-scale dynamic ride-sharing for taxis and evaluate it based on Beijing trips.
\citet{Zhang2014carpool} present a scalable framework of ride sharing with a focus for carpool capabilities (i.e. a taxicab with a passenger and a route, picking up another passenger for a similar trip) and evaluate it on trips in the city of Shenzhen. They show that their system, \textit{CallCab}, can reduce the total traveled distances by 60\% and wait times by 41\%. 
\citet{khan2017ride} posit that ride-sharing is more effective when users can go out of their way slightly and also agree on common destinations.

A less studied scenario of shared transportation is car sharing. 
\citet{Katzev2003} studied effects of car sharing on mobility and environmental impacts. 
Several similar, recent studies have investigated mobility and impacts of car sharing \cite{Baptista2014,Nijland2017,Musso2012}.
\citet{HANNA2016254} formulated the car sharing problem as a bipartite graph and solved the matching for three criteria: min cost, min makespan (limit of distance for matches), and strategic manipulability (no user has the incentive to misreport its location and hence distance).
In \cite{agatz2016autonomous}, authors discuss several challenges of realizing car and ride sharing, including optimization of multiple trade-offs (number of vehicles, convenience, total travel time, operating costs).
This new concept of car sharing and ownership, coupled with the rise of autonomous and automated transportation systems (e.g. self-driving cars) might change the face of mobility in urban environments. In this paper, we present a novel formulation for scheduling a car sharing/automated fleet of vehicles with the minimum number of cars.
\begin{figure}[!t]
	\subcaptionbox{8-9am distance}{%
		\includegraphics[width=0.49\columnwidth]{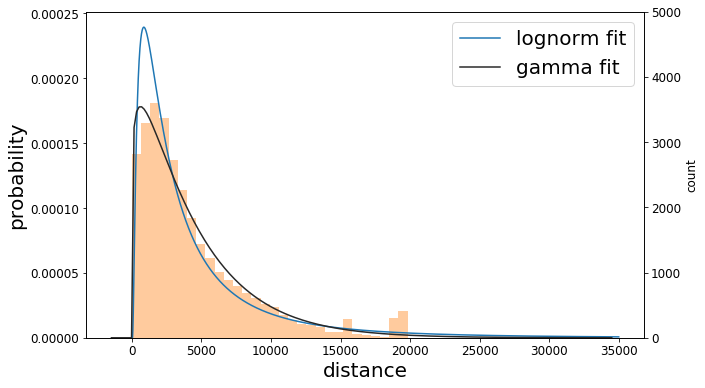}%
    }
	\subcaptionbox{4-5pm distance}{%
		\includegraphics[width=0.49\columnwidth]{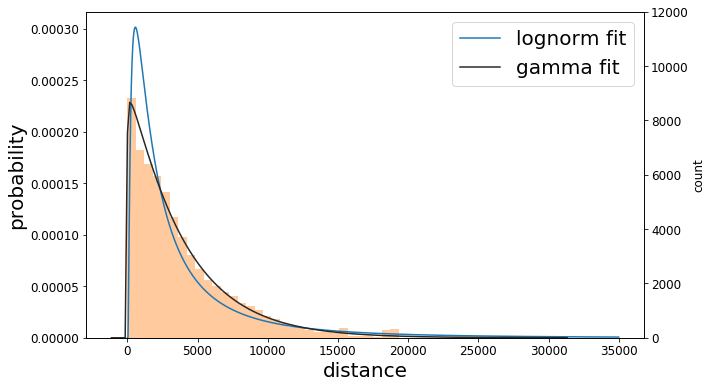}%
    }
	\subcaptionbox{8-9am duration}{%
		\includegraphics[width=0.49\columnwidth]{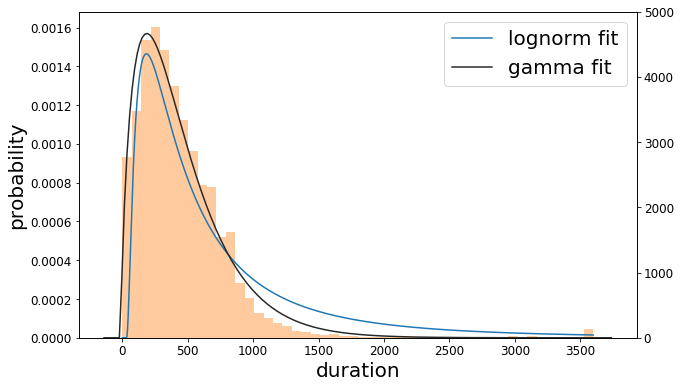}%
    }
	\subcaptionbox{4-5pm duration}{%
		\includegraphics[width=0.49\columnwidth]{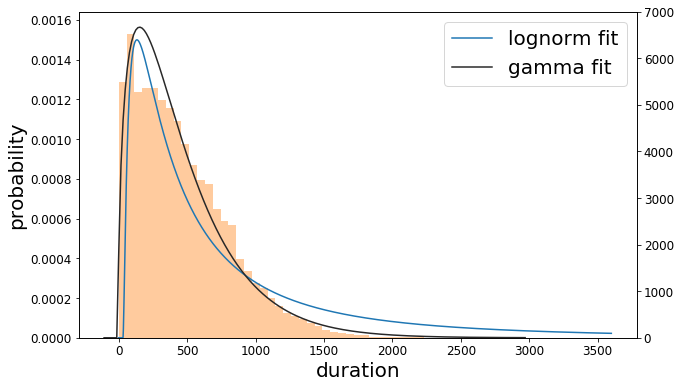}%
    }%
    \caption{Distribution of duration and traveled distance of trips in different times of day.}
    \label{fig:distr_dist_dur}
\end{figure}

\section{Dataset and Analysis}\label{sec:analysis}

The dataset used for this study is mobility traces for the city of Cologne \cite{kolntmc}.
The data was collected as part of an initiative to realistically reproduce vehicular mobility in the city\footnote{http://kolntrace.project.citi-lab.fr/}. We choose this dataset as it provides us with fine-grained traces of trips in a long timespan of 24hours in a major city.
This dataset contains traces consisting of location, time and speed of vehicles on a trip (700K trips, 350M records in a 400 $km^2$ area). To analyze different times of the day and understand temporal dynamics, we select four 1-hour time windows throughout the day (Morning 8-9am, Noon 12-1pm, Evening 4-5pm, Night 8-9pm). For the OD representation of the trips, the first and last appearances of each ID are extracted.

\subsection{Exploratory Analysis of Spatial and Temporal Characteristics}
We start with an exploratory analysis of the dataset in order to better understand the vehicular mobility through the characteristics that pertain to the trips. We comparatively investigate (different times of the day) the trip duration and distance distributions in \ref{sssec:distr} (more accurately the distance between origin and destination represents the displacement and we use them interchangeably), number of trip waypoints (records in data) in \ref{sssec:wp}, spatial spread of the trips in \ref{sssec:spatial}, temporal distributions (in \ref{sssec:temporal}) and the relation of trip duration to trip start location in \ref{sssec:relation}.

\subsubsection{Duration and Distance Distribution}\label{sssec:distr}
\begin{figure}
\setlength{\belowcaptionskip}{-13pt}
	\subcaptionbox{} {\includegraphics[width=0.49\columnwidth]{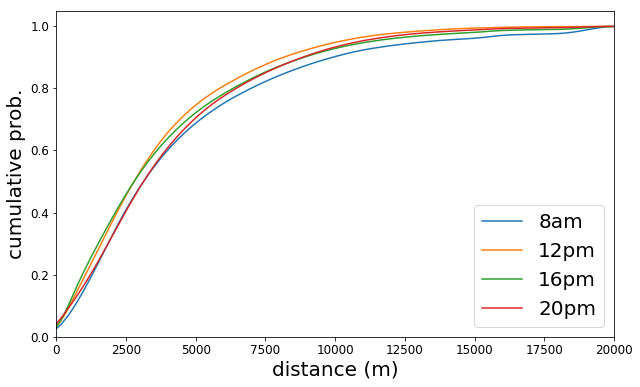}}
    \subcaptionbox{} {\includegraphics[width=0.49\columnwidth]{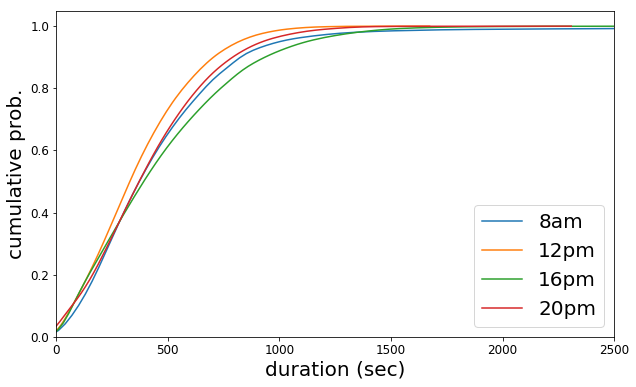}}
    \caption{Comparison of trip (a) distance (b) duration through out the day.}
    \label{fig:cdf_dist_dur}
\end{figure}

Figure \ref{fig:distr_dist_dur} shows the distribution of trip duration and distance in each hour of the day. As observable, the distribution in all cases is right-skewed. Since one common distribution to model the trip duration is log-normal \cite{enroute,dandy1984variability,rakha2010trip,arroyo2004modeling}. The fit of lognormal is plotted on the histograms and compared to Gamma distribution which provides a better fit (visually) as it is a more flexible (general) distribution. The duration has a lighter tail comparing to lognormal. Figure \ref{fig:cdf_dist_dur} compares duration and traveled distance in different times of day. Trips that happen around evening rush hour are longer than rest of the day (e.g., $\approx90\%$ of trips are at most $\approx 1000$ seconds vs $\approx 700$ seconds for noon). As for the distance, similar trait holds except the probability of shorter (3KM or less) trips are almost the same. Also, morning trips tend to be lengthier than other times of the day.

\begin{figure}[b]
	\includegraphics[width=0.66\columnwidth,height=0.4\columnwidth]{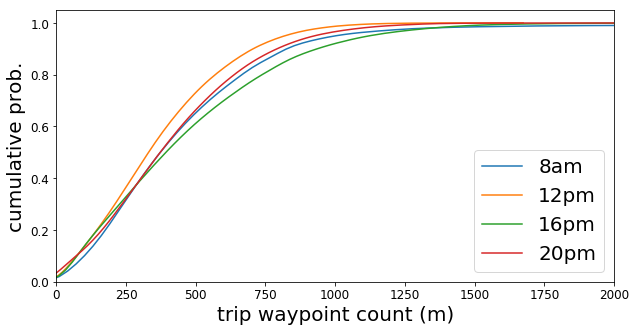}
    \caption{CDF of number of waypoints per trip in different times of day.}
    \label{fig:cdf_waypoint}
\end{figure}

\begin{figure*}[!ht]
	\subcaptionbox{8-9am\label{fig:distance_distr:a}}{%
		\includegraphics[width=0.49\columnwidth]{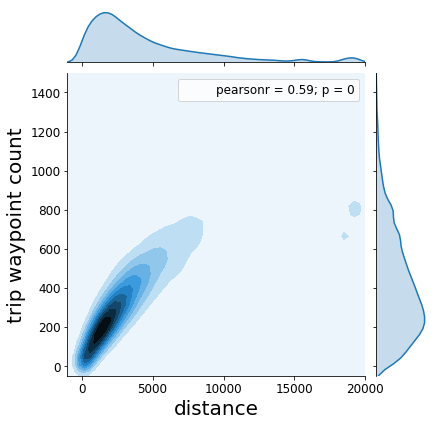}%
    }
	\subcaptionbox{12-13pm\label{fig:distance_distr:b}}{%
		\includegraphics[width=0.49\columnwidth]{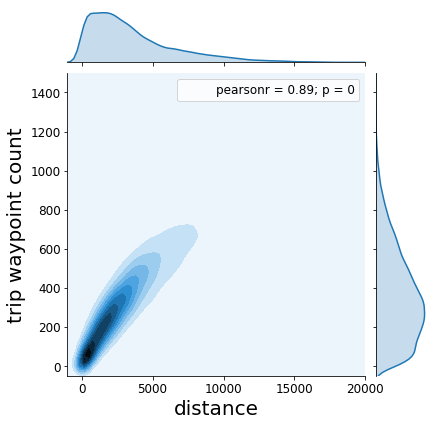}%
    }
	\subcaptionbox{4-5pm\label{fig:distance_distr:c}}{%
		\includegraphics[width=0.49\columnwidth]{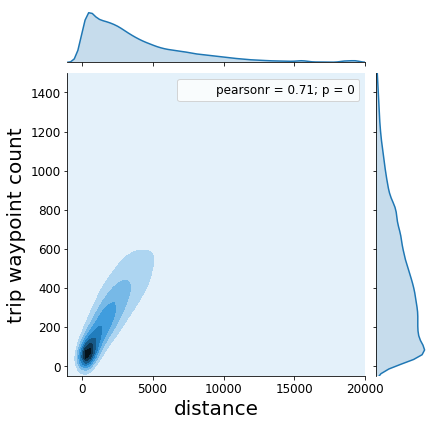}%
    }
	\subcaptionbox{8-9pm\label{fig:distance_distr:d}}{%
		\includegraphics[width=0.49\columnwidth]{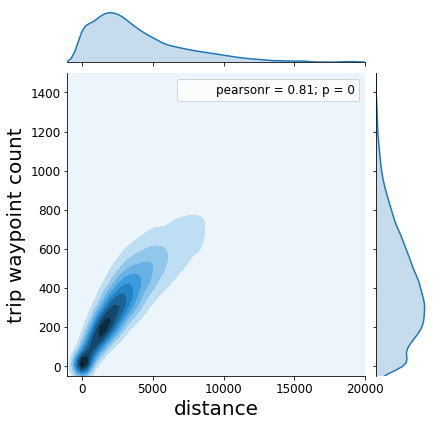}%
    }
    \caption{Joint distribution of count of waypoints for a trip vs distance in different times of day.}
    \label{fig:joint_plot_waypoint_distance}
\end{figure*}

\begin{figure*}[!ht]
	\subcaptionbox{8-9am}{%
		\includegraphics[width=0.48\columnwidth]{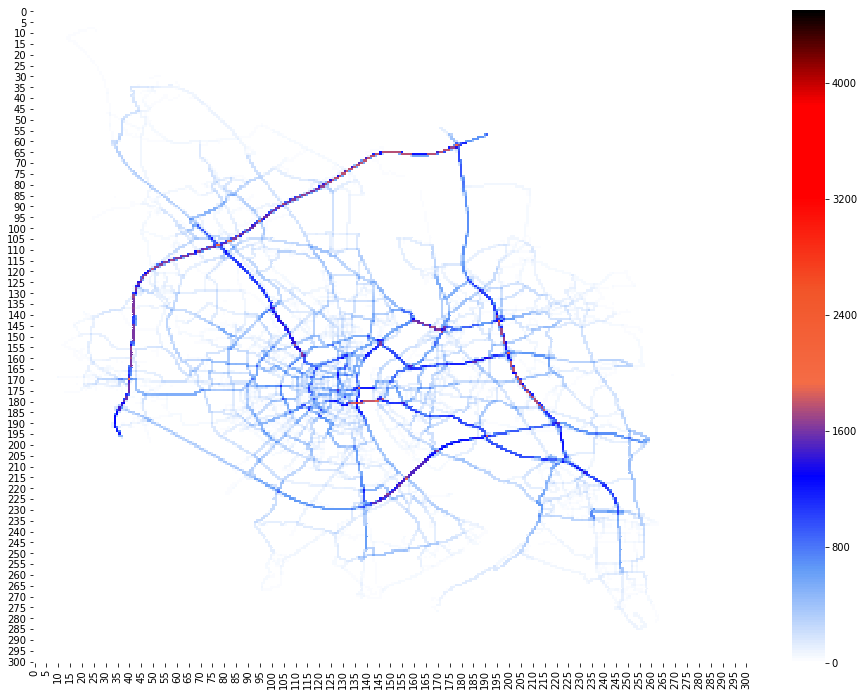}%
    }
	\subcaptionbox{12-13pm}{%
		\includegraphics[width=0.48\columnwidth]{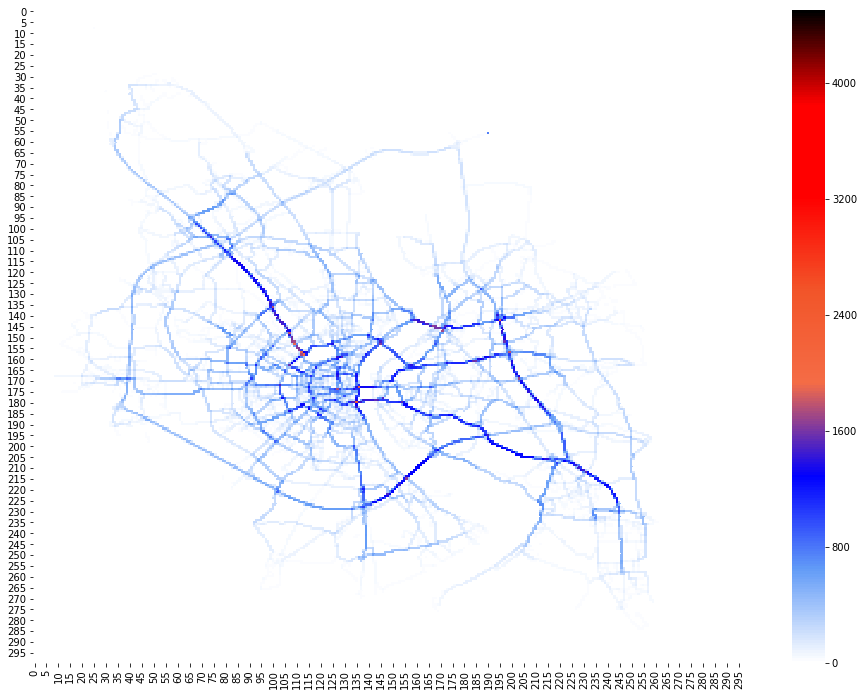}%
    }
	\subcaptionbox{4-5pm}{%
		\includegraphics[width=0.48\columnwidth]{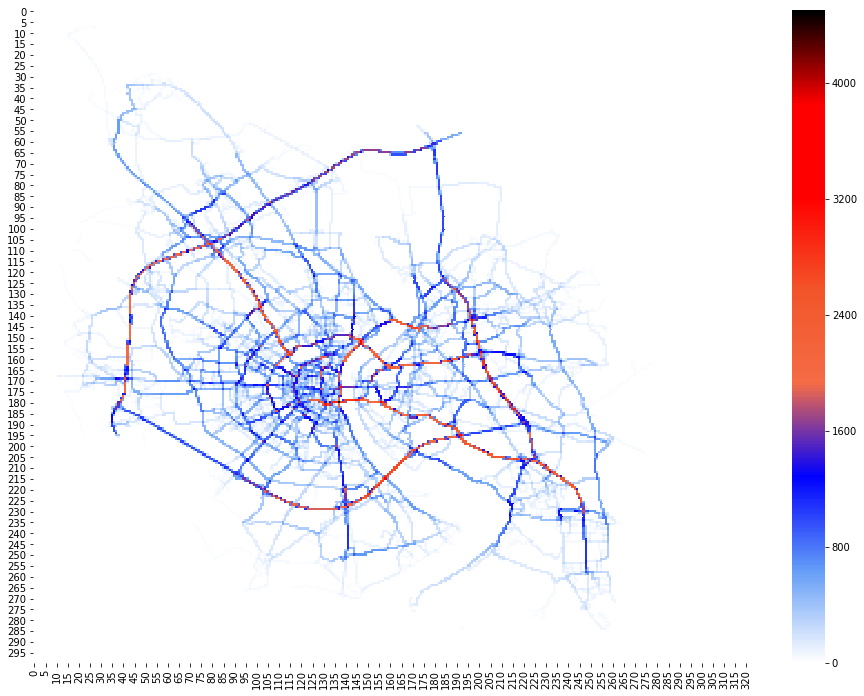}%
    }
	\subcaptionbox{8-9pm}{%
		\includegraphics[width=0.48\columnwidth]{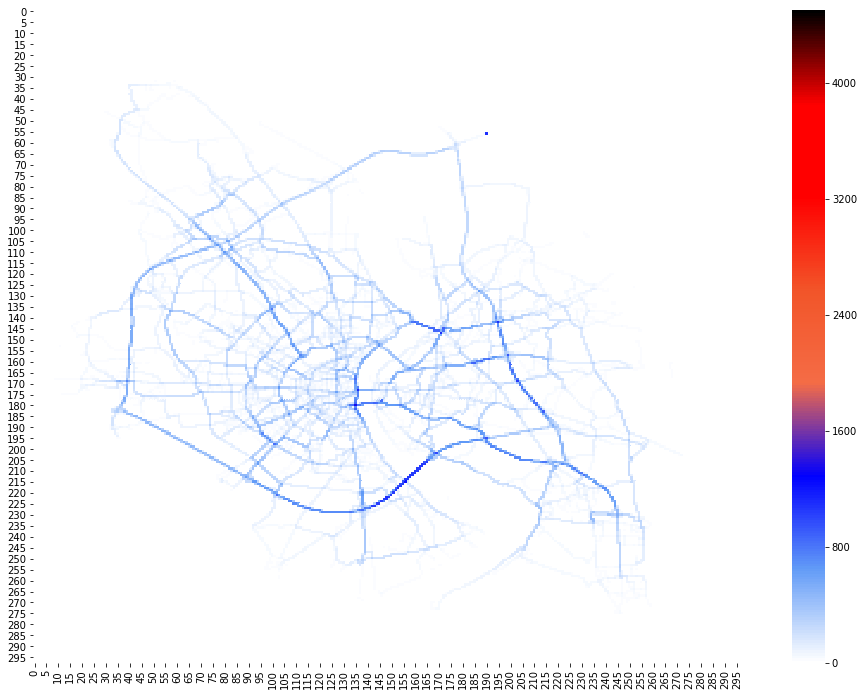}%
    }
    \caption{Heatmap of unique number of trips spanning the city (on a 300x300 grid). It resembles the city map with its traffic load in different times of the day.}
    \label{fig:heatmaps_uniq}
\end{figure*}

\subsubsection{Number of waypoints}\label{sssec:wp}
Waypoints are simply put, reappearances of the same trip\_id in traces. Although this measure is not an intrinsic characteristic of the trips, rather the result of data generation method, it can provide us with insights for understanding the dataset and later for feature engineering and representations for the trips.

Figure \ref{fig:cdf_waypoint} shows the cumulative distribution of the number of waypoints per trip throughout the day. The waypoint count is higher for 4 pm trips ($\approx90\%$ of having roughly 1000 waypoints or less, while at noon the number becomes $\approx$750). Morning and night periods (8 am and 8 pm) follow similar patterns. To further address the implications of such data we compare waypoint count to distance. Figure \ref{fig:joint_plot_waypoint_distance} presents  joint-distribution plots of waypoint count and traveled distances. The Pearson correlation coefficient is 0.59, 0.89, 0.71 and 0.81 for the hours of the day in the study respectively which suggests the existence of a relatively strong positive linear correlation between the two peaking at noon time. This, in turn, suggests that the number of waypoints can be a good representative of the distance traveled in this data (possibly due to equidistant time intervals of trip logging in data generation).

  \subsubsection{Spatial Spread}\label{sssec:spatial}
  Figure \ref{fig:heatmaps_uniq} shows 300x300 heatmaps, depicting the structure of the city based on the volume of traffic (unique count of trips in a grid cell in the hour).
  At noon, the traffic seems to be more focused on the middle and south-east side of the city with less load on the highways/beltways.
  There is a significant drop in traffic towards the night. The spatial spread of the trips represents the map of city roads, this suggests the existence of a subspace in the spatial space of trips (so a more compact representation might be more useful comparing to lat-lon. coordinate space).

\begin{figure}[t]
	\subcaptionbox{} {\includegraphics[width=0.49\columnwidth]{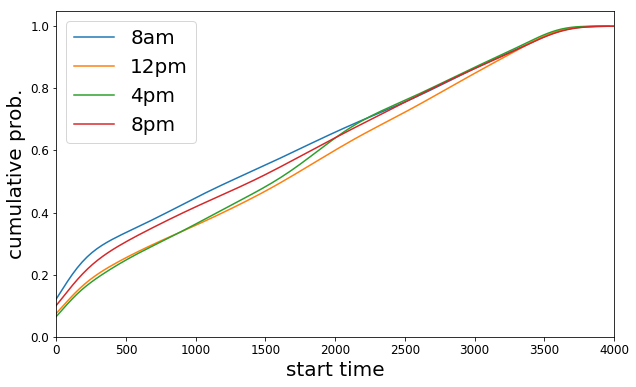}}
    \subcaptionbox{} {\includegraphics[width=0.49\columnwidth]{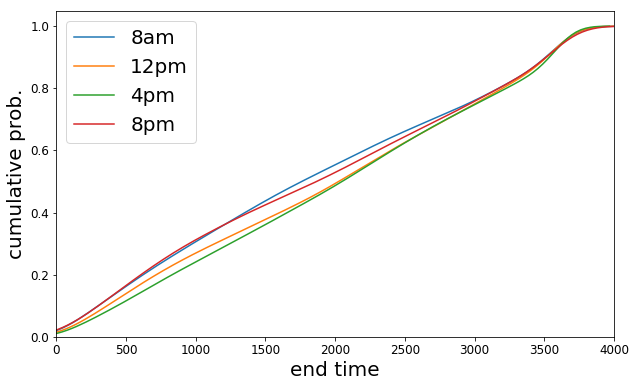}}
    \caption{Comparison of trip (a) start time (b) end time through out the day.}
    \label{fig:cdf_temporal}
\end{figure}

\subsubsection{Temporal Distribution}\label{sssec:temporal}
CDFs of start and end times of the trips are shown in figure \ref{fig:cdf_temporal}. More morning and night trips start in the first 10 minutes of the hour than the first 10 minutes of noon or evening rush hour although in the case of 4 pm, towards the end of the hour, more trips are started (there is a jump in the number of trips that start after 2200 seconds). At half hour marker, almost 10\% more trips have already finished in the morning and nighttime comparing to noon or 4 pm (close to 50\% vs 40\%).

\subsubsection{Relation of space and time}\label{sssec:relation}
\begin{figure}[t]
\setlength{\belowcaptionskip}{-13pt}
    \includegraphics[width=0.99\columnwidth]{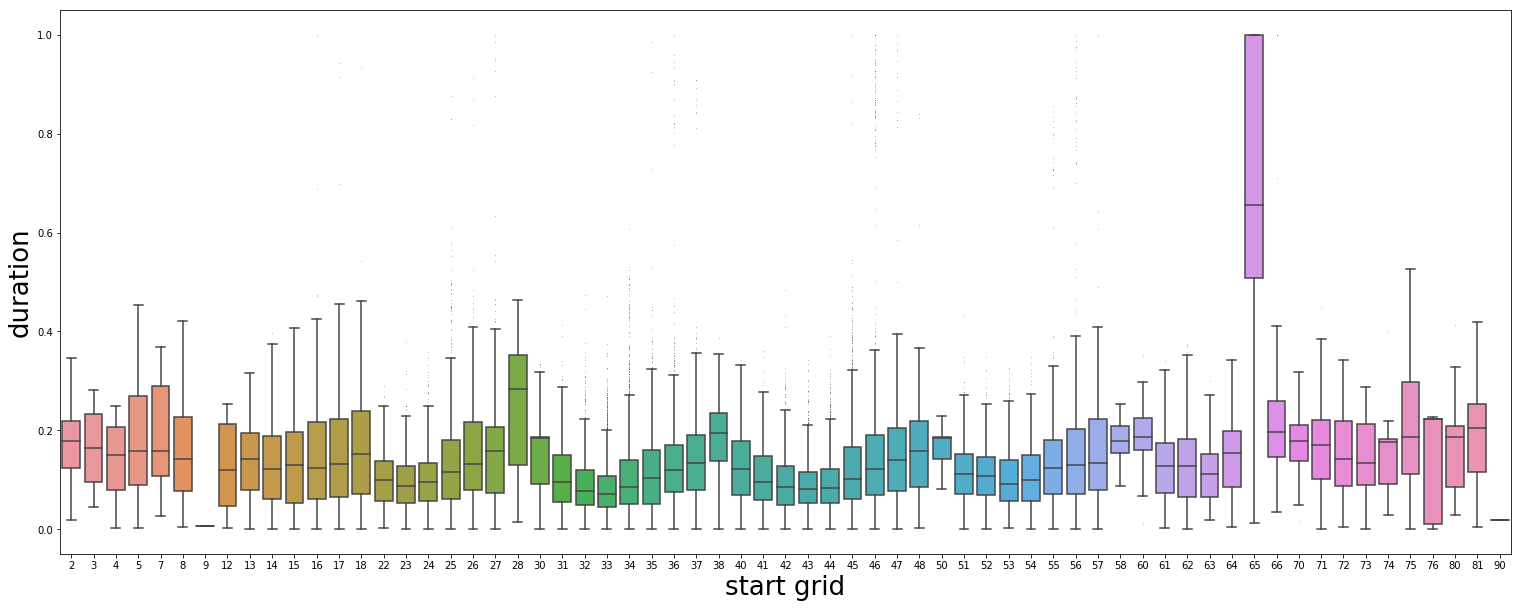}
    \caption{Box plot of duration of 8am trips versus their starting cell in a 10x10 grid of the city area. Repeating patterns are observed on west to east traversal of the grid. Overall trend shows a slight increase in duration towards the north.}
    \label{fig:box_plots}    
\end{figure}

We observe the relationship of trip start location and its duration using box plots in Fig. \ref{fig:box_plots} throughout the day. The repeating patterns each corresponds to same longitude grid cells (left to right, going from west to east) and the overall trend moving from left to right in the figures corresponds to higher latitudes. The patterns suggest that trips happening in the east of the city are longer in general and trips happening on the north side are slightly longer as well. This may be explained by highways (German Autobahn) around the city to the east and north-west.

\subsection{Similarity}\label{sec:similarity}
In order to better understand the trips, we propose a similarity measure as the arithmetic mean of point-wise (e.g. origin vs origin, destination vs destination) distances each achieved through the weighted geometric mean of Euclidean similarity (spatial) and their temporal similarity. Both location and time features are scaled to range $[0,1]$.
The intuition behind such a choice is that the Geometric mean adjusts the spatial component of two points based on their temporal similarity and arithmetic mean provides a measure based on all the point-wise similarities. Mathematically:

$$ psim(p_1,p_2) = e^{\frac{w_1 \ln(\frac{1}{1+dist(p_1,p_2)}) + w_2 \ln(\frac{1}{1+time(p_1,p_2)})}{w_1+w_2}}$$
$$ sim(t_1,t_2) = \frac{\sum_{i=0}^n psim(t_1^i,t_2^i)}{n} \label{math:sim_measure}$$

Where $dist(p1,p2)$ is the spatial distance of two points (Euclidean in this case), $time(p_1,p_2)$ is the absolute difference of the points in time and $w_1$ and $w_2$ are the weight given to them respectively. We use $1/(1+dist)$ to convert a distance into similarity. The similarity of trips $sim(t_1,t_2)$ is the average of all the $n$ points (waypoints, or just endpoints in case of OD representation). In this study, for clustering and matching of trips, higher weight ($w_1 = [0.6 , 0.7]$ vs $w_2 = [0.3 , 0.4]$) is given to the spatial component so that time difference acts as the adjusting factor in determining the similarity measure for a pair of trips. Effect of the choice of weights on the results of the matching is available in Sec. \ref{sec:compare}.

\begin{figure}[t]
\setlength{\belowcaptionskip}{-13pt}
    \includegraphics[width=0.55\columnwidth]{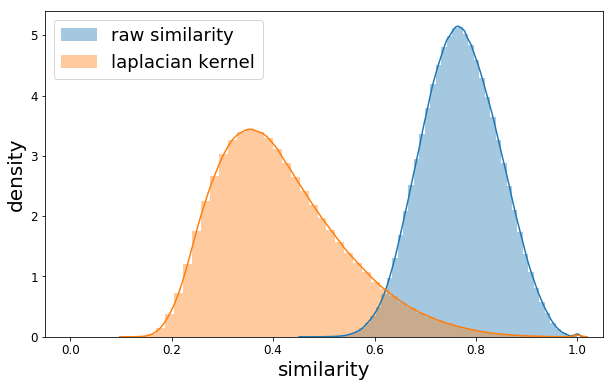}
    \label{fig:similarity measure distribution}
    \caption{Distribution of the values of similarity. Laplacian kernel is applied to create more distinction between values. }
\end{figure}

\subsection{Clustering and Visualization}\label{sec:clustering}
With the help of similarity defined above in section \ref{sec:similarity}, we are able to generate spatio-temporal clusters by using Spectral Clustering (SC). Because of the spatial constraints enforced by the map of the city, spatial features do not form a convex space and hence we need clustering algorithms that work with connectivity instead of shape. We tried DBSCAN and SC algorithms; Only SC was able to find meaningful clusters. SC make use of the spectrum (eigenvalues) of the similarity matrix of the data to project the data into fewer dimensions before clustering.

\begin{figure}[t]
\setlength{\belowcaptionskip}{-5pt}
	\subcaptionbox{} {\includegraphics[width=0.49\columnwidth]{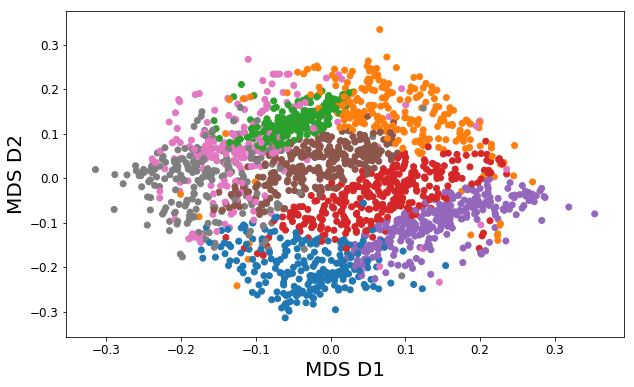}}
    \subcaptionbox{} {\includegraphics[width=0.49\columnwidth]{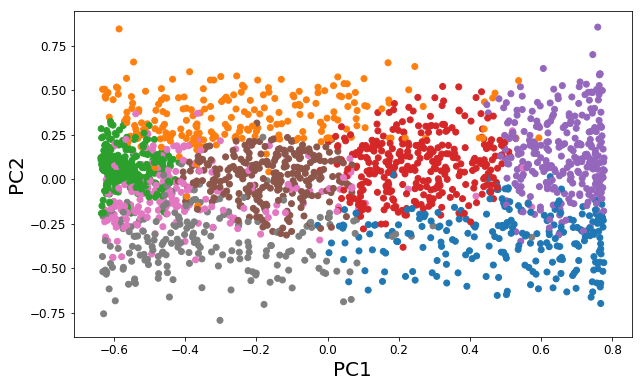}}
    \caption{Results of Spectral Clustering for 8-9am visualized (each color represents a cluster) on (a) 2D MDS plot (b) First two Principal Components. Other times of the day exhibit similar patterns.}
    \label{fig:clustering_mds_pca}
\end{figure}

\begin{figure}[t]
\setlength{\belowcaptionskip}{-5pt}
	\includegraphics[width=0.55\columnwidth]{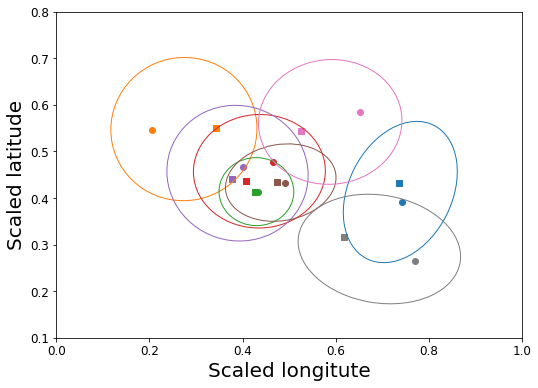}
    \caption{Spread of clusters in space}
    \label{fig:clustering_spatial}
\end{figure}

\begin{figure}[!h]
\setlength{\belowcaptionskip}{-13pt}
	\subcaptionbox{} {\includegraphics[width=0.49\columnwidth]{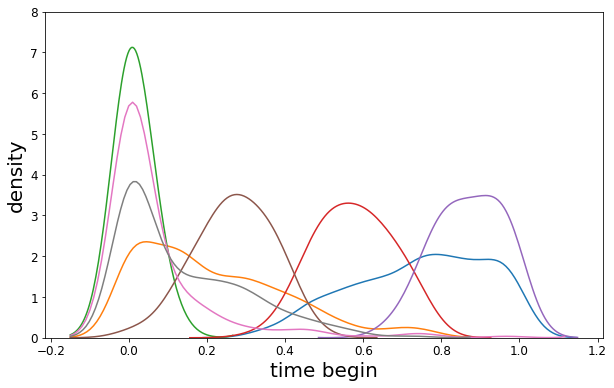}}
    \subcaptionbox{} {\includegraphics[width=0.49\columnwidth]{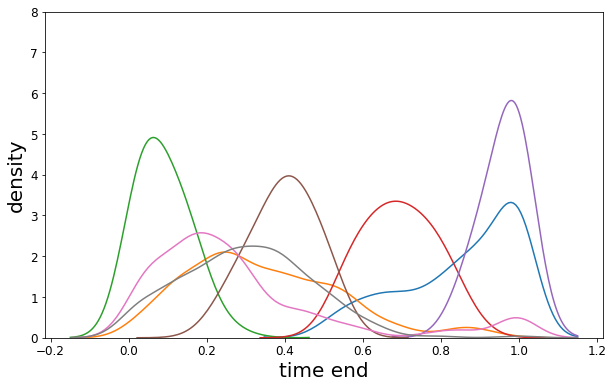}}
    \caption{Kernel Density Estimates of 8am trips (a) start times (b) end times. There is a clear separation in time for the clusters that are spatially intermixed.}
    \label{fig:clustering_time}
\end{figure}

In order to visualize the data, Multi-Dimensional Scaling (MDS) and Principal Component Analysis (PCA) are used. Figure \ref{fig:clustering_mds_pca} shows the resulting clusters for morning trips on 2D scaling and 2D projection of OD representation into first two principal component's space. First two PCs capture over 82\% of the variance of the data. 
To test the quality of this clustering, we attempt to interpret the resulting clusters.
Clusters are shown on a scaled lat. lon. plane with mean and median of start and end points in Figure \ref{fig:clustering_spatial}. The ellipsoid axes correspond to the std. of lat. and lon. of the cluster. These clusters cover the area of the city with some of them spanning same areas especially near the city's center (downtown). Interestingly, as Figure \ref{fig:clustering_time} shows, there is a clear separation in time for those clusters that are spatially overlapping. This suggests the clustering scheme is successful in finding clusters that show different spatio-temporal traits.

\section{Matching Trips with OD Representation}\label{sec:matching}
The existence of clusters and a spectrum of similarity suggest the availability of trips nearby.
Knowing this we can use variations of the similarity score for  applications such as shared transportation.
Thus, we refine the similarity to reflect more on each scenario as well as a dynamic matching of trips for each case.
We focus on morning time window (8 am) and for all cases, the trips are divided into two categories: Riders and Rides (2000 riders vs 10000 rides). The goal is to find matches for the rider to ride with. Trip end times are used as the expected time of arrival (latest possible).

\subsection{Catch-a-Ride}
In this scenario, riders have a time margin to reach the pickup points (ride's origin) and a margin for them to arrive at their destination after drop-off (ride's destination). We redefine the similarity score as Catch-a-Ride (CaR) score between $t_1$ and $t_2$ as if $t_1$ can move to $t_2$ and ride with it. Therefore, $t_2$ has to start after $t_1$ and finish before it to have a high score (and be close nearby). We can do this by modifying the similarity score so that the time component is the time of $t_2 - t_1$ for the origin points and time of $t_1 - t_2$ for destinations rather than the absolute values. This will create an asymmetric affinity matrix $A$ where $A_{i,j}$ is the $CaR\_score(t_i,t_j)$.

\subsubsection{Clustering and Visualization} 
\begin{figure}[!t]
\setlength{\belowcaptionskip}{-13pt}
	\subcaptionbox{} {\includegraphics[width=0.49\columnwidth]{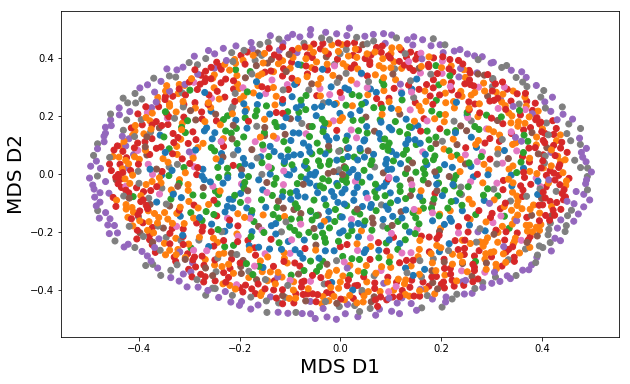}}
    \subcaptionbox{} {\includegraphics[width=0.49\columnwidth]{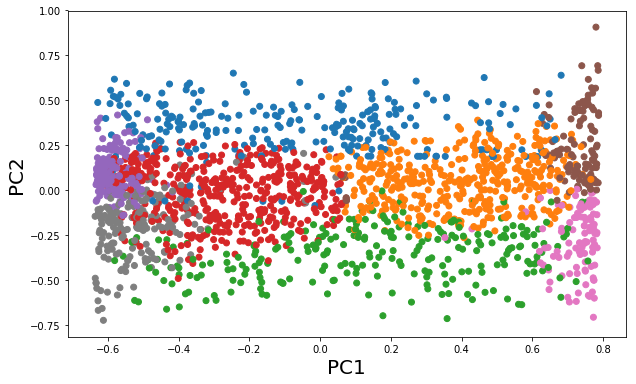}}
    \caption[Visualization of Catch-a-ride clusters]{Visualization of the clusters on (a) 2D DMS (b) First 2 PCs. CaR score results in nested ellipses in MDS. By definition, $CaR\_score$ and $carpool\_score$ are complements of each other with respect to original similarity score and hence the clusters (and MDS and PCA) would be the same (with some rotation).}
    \label{fig:clustering_CaR}
\end{figure}

We decompose the matrix into one symmetric (average of the matrix and its transpose) and one anti-symmetric (skew-symmetric) component (half of the difference of matrix and its transpose) and then use the symmetric part for clustering and visualization. Fortunately, the ratio sum of squares of the symmetric matrix to the original is 98\% which suggests the matrix is almost symmetric in nature and the symmetric component can be a good representative of it. This is expected as the CaR\_score is a subset of the similarity score where the time difference is not the absolute value. Clusters (using spectral clustering) are visualized in Figure \ref{fig:clustering_CaR} on MDS and 2PC plots.

\subsubsection{Dynamic Matching}
\begin{figure}[t]
	\centering
    \setlength{\belowcaptionskip}{-10pt}
    \subcaptionbox{} {\includegraphics[width=0.49\columnwidth]{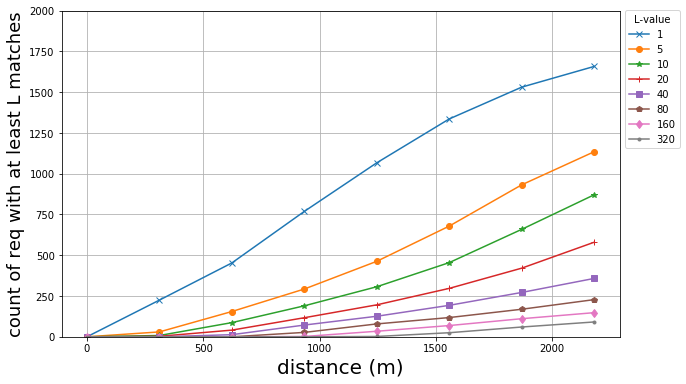}}
    \subcaptionbox{} {\includegraphics[width=0.49\columnwidth]{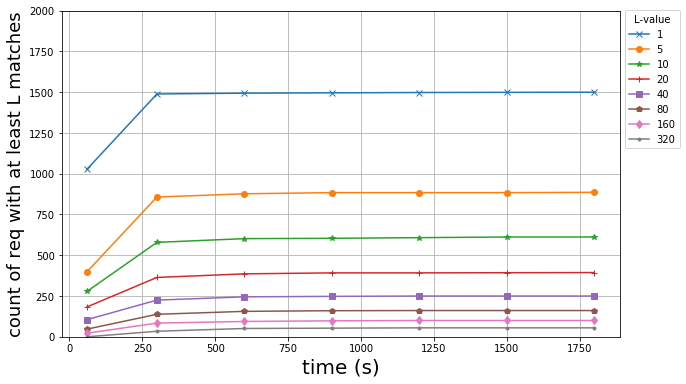}}
    \caption{Number of potential matches for CaR (similarly for CarPool) (a) per distance (time threshold fixed at 900 sec) (b) per time (distance threshold fixed at 1800 meters).}
    \label{fig:CaR_count_per_dist_and_time}
\end{figure}

For each rider, the potential matches are filtered out by using distance and time thresholds. The trend of the number of matches per threshold is presented in Figure \ref{fig:CaR_count_per_dist_and_time} (the number of riders where they have at least L matches). As observable, the number of matches is more sensitive to distance thresholds compared to time. After 250 seconds, the number of riders that get a match at any value of L scantly changes. At any distance threshold, the number of riders with at least L matches significantly drop with an increase of L ($\approx$900 riders get at least 5 matches while $\approx$1500 get at least a match). Next, we choose the thresholds of 1800 meters and 900 seconds (which enables us to provide a match for almost 3/4 of the riders) and use the CaR\_score as the matching criteria. Note that the choice of one is independent of others and hence the greedy approach is optimal.
The results of such a matching scheme are presented in Table \ref{tab:dynamic_match_car_carpool}. The total travel required by both the requests (riders) and the matches (rides) are $12990.199$ KM where 33.5\% of it is traveled by the matches.
Since roughly 75\% of the requests have a match (1496 out of 2000), the ratio of matched to total for those is 45.4\% which means in cases where at least a potential match exists, \textbf{$\approx$55\%} of the traveled kilometers can be saved.  
Since 5235.3 km of the 8633.8 are taken by those who have matches, the total travel done by unmatched requests is 3398.5 KM. Then, the traveled distances with ride-sharing amount to 3398.5 + 4356.3, the ratio of which to total (12990.2) is 59.7\% which yields \textbf{a saving of 40.3\%}, compared to having no shared transportation at the cost of commuting to pick up and drop off locations ($\approx$1000 KM each, average of 0.66 KM for each rider).

\newcolumntype{b}{>{\footnotesize}X}
\newcolumntype{s}{>{\hsize=.33\hsize}X}

\begin{table*}[ht]
\centering
\caption{Results of matching trips for ride sharing Catch-a-Ride (CaR) and carpool (CP) scenarios.}
\label{tab:dynamic_match_car_carpool}
\begin{tabularx}{\textwidth}{bss}
\hline
      & CaR & CarPool \\ \hline
match travels (km) & 4356.368               &  5486.785          \\
req travels (km) & 8633.831               &   8633.831 \\
match to total travel ratio & 33.5\% & 38.8\%\\
origin-origin distance (km) & 1017.665 & 948.438 \\
dest-dest distance (km) & 1045.912 & 1019.560\\
origin-origin times (sec) & 70185 &  83171\\
dest-dest times (sec) & 88873 & 88064 \\
\# req with at least a match & 1496 & 1458\\
req travels for least a match (km) & 5235.319 & 4938.073 \\
match to total travel ratio (at least a match) & 45.4\% & 52.6\% \\
\hline
\end{tabularx}
\end{table*}

\subsection{Car Pooling}
Similar to the previous scenario, but instead the ride is willing to go out of their way (within a threshold) to pick the rider up and drop them off before arriving at its own destination on time. The clustering is omitted as the constructed affinity matrix is the transpose of CaR scenario and hence would result in the same symmetric component. The dynamic matching is discussed next. The threshold filters are applied to find potential matches.

\begin{figure}
\centering
	\subcaptionbox{}{\includegraphics[width=0.49\columnwidth]{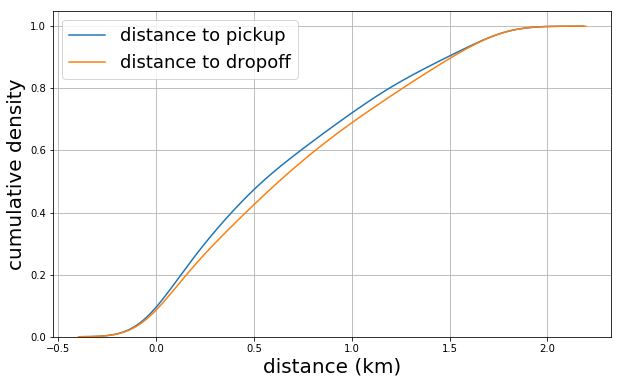}}
    \subcaptionbox{}{\includegraphics[width=0.49\columnwidth]{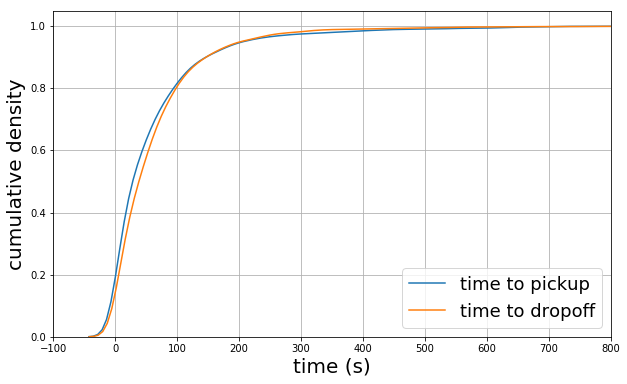}}
    \caption{CDF of CP results of pick-up and drop-off (a) distance (b) time.}
    \label{fig:CP_result_dist_time}
    \vspace{-10pt}
\end{figure}

The trends of the number of potential matches over time and distance thresholds follow a very similar curve to CaR (Fig. \ref{fig:CaR_count_per_dist_and_time}). There is a slightly lower number of requests with a match at each threshold, but the overall trends are alike with more sensitivity to distance thresholds. Distance and time thresholds of 1800 meters and 900 seconds are again applied, and the CP\_score is used as the matching criteria.
Results are presented in Table \ref{tab:dynamic_match_car_carpool}. The total KMs traveled would be the request travels plus the pickup and drop off ($8633.8 + 948.4 + 1019.5 = 10601.7$). Without any carpooling, the total travels by matches and requests would be $8633.8 + 5486.8 = 14120.6$KM and hence in existence of the ride-sharing there is 25\% decrease in total KMs traveled at the cost of drivers going a total of 948.4 KMs out of their way (average of 0.65 KM per driver) for pick up and 1019.5 KMs for drop off (0.69 KM average). CDFs of resulting pick-up and drop-off distances and times are presented in Figure \ref{fig:CP_result_dist_time}.

\section{Comparison with Other Similarity Scores}
\label{sec:compare}
\begin{figure}
\centering
\setlength{\belowcaptionskip}{-10pt}
	\subcaptionbox{}{\includegraphics[width=0.49\columnwidth]{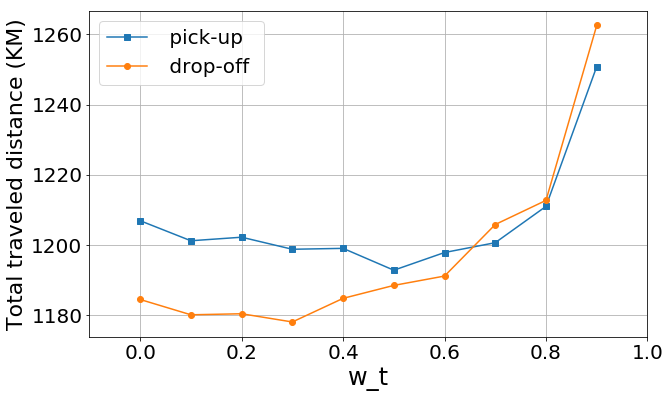}}
    \subcaptionbox{}{\includegraphics[width=0.49\columnwidth]{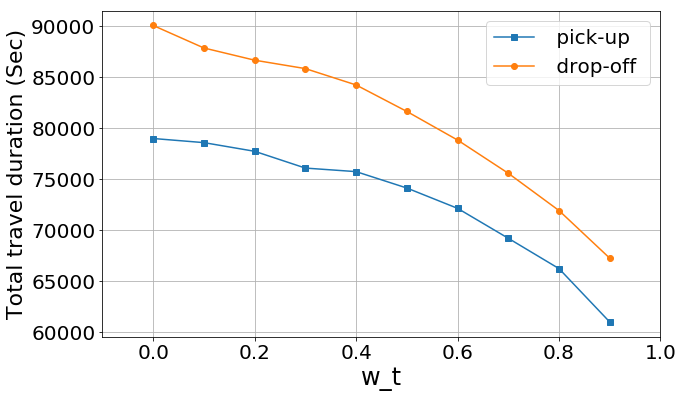}}
    \caption{pick-up and drop-off (a) distance (b) time.}
    \label{fig:oodd_dist_time_per_wt}
\end{figure}

The explained scenarios only utilize the OD representation for the trips (and hence only the endpoints) but the similarity score is not limited to OD representation. As mentioned in the \ref{sec:relatedwork} section, various measures of similarity or distance have been proposed for comparison of trajectories. Longest Common Sub-Sequence (LCSS) is a measure of closeness based on the number of points shared by two trajectories (points within thresholds of each other). Frechet distance is the maximum distance between two curves over all parameterization of the curves. This can be imagined as the minimum length of the leash required to take a dog for a walk when the human and dog have their own path to travel with all the possible combinations of speed for each (no backward movement allowed). Dynamic Time Warping (DTW) works by finding the minimum cost warp path where a warp path is the sequence of changes that warps a trajectory into the other. 
In this section, we investigate the performance of Catch-a-Ride scenario using DTW, Frechet Distance and LCSS and our score (referred to as WGM for Weighted Geometric Means, weights of 0.6 and 0.4 for distance and time resp.). In all cases, we use a 50-waypoint (sampled) representation of the trips instead of only OD and the same filtering is applied (and hence measures related to the request set are kept constant). The results are presented in table \ref{tab:comparison}. In case of WGM with time, the weights of the geometric mean are given as 0.1 for spatial and 0.9 for temporal parts. In DTW with time, the cost is the product of distance and time difference instead of distance only. As the table shows, Frechet distance (which is an upper-bound on the distance of the trajectories by definition) provides closest matches in distance (but not in time). On the other hand, DTW with time, results in least orig-orig and dest-dest time but \textbf{not} distance. WGM achieves the results with both time and space in mind so at a tiny cost in orig-orig and dest-dest distance it beats others (original versions) in temporal aspect . Also with the flexibility of weights, we can tune it for more focus on time or space. From a runtime point of view, our score is linear in the number of waypoints for the trips whereas LCSS and DTW have runtime and space complexity of $O(N^2)$ and Frechet takes even longer with $O(N^2 log(N))$. To better understand the sensitivity of results to the choice of the weight of the temporal component in WGM, trends of change in pick-up and drop-off distance and time is plotted against temporal weight (w\_t) in Fig. \ref{fig:oodd_dist_time_per_wt}. With an increase in w\_t, more preference is given to time element of the similarity and hence the matches are closer in time (total pick-up and drop-off times decreases) while they are not necessarily closer in space. This can be used to tune the score for different applications with levels of dependence on time versus location.

\begin{table*}[ht]
\centering
\caption{Comparison on different similarity/distance scores on Catch-a-Ride scenario.}
\label{tab:comparison}
\begin{tabularx}{\textwidth}{bssssss}
\hline
      & WGM & LCSS & Frechet & DTW & DTW w/ time & WGM w/ time\\ \hline
match travels (km) & 5655.997  &  5393.861 & 5671.631 & 5652.613 & 6038.372  & 5945.291  \\
req travels (km) & 11706.334   & 11706.334 & 11706.334 & 11706.334 & 11706.334 & 11706.334 \\
match to total travel ratio & 32.57\% & 31.54\% & 32.63\% & 32.56\% & 34.02\% & 33.68\% \\
origin-origin distance (km) & 1149.131 & 1442.782 & 1104.907 & 1127.623 & 1202.733 & 1203.724 \\
dest-dest distance (km) & 1188.705 & 1421.046 & 1144.947 & 1165.916 & 1222.000 & 1255.597 \\
origin-origin times (sec) & 71365 & 86071 & 74544 & 72567 & 52866 & 57896\\
dest-dest times (sec) & 80860 & 101652 & 89570 & 86655 & 57546 & 62424 \\
\# req with at least a match & 1357 & 1357 & 1357 & 1357 & 1357 & 1357\\
req travels for least a match (km) & 7244.179 & 7244.179 & 7244.179 & 7244.179 & 7244.179 & 7244.179 \\
match to total travel ratio (at least a match) & 43.84\% & 42.67\% & 43.91\% & 43.82\% & 45.46\% & 45.07\% \\
\hline
\end{tabularx}
\end{table*}

\section{Car Sharing}\label{sec:carsharing}
The similarity score, in addition to clustering or dynamic matching of trips, can be used to find a scheduling for car sharing applications. This is particularly useful in case of the existence of autonomous driving systems. Our aim here is to service all our requests (trips) with the minimum number of cars shared while in all the possible assignment, we focus on the ones that minimize the distance and time different (based on the similarity score) between consecutive trips assigned to the same car. The same set of 2000 trips is used as the request set. The problem is formulated as a graph:
\begin{itemize}
\item Each node represents a trip (N nodes). Each directed edge between any 2 given nodes represents the possibility of the destination node to use the car/get serviced after the source node of that edge (i.e. an edge $(a,b)$ exist only if the end point of $a$ is within time and distance threshold of start point of $b$ and $b$ starts after $a$ ends). The similarity score is used as edge weights.

\item The problem as stated above (i.e. finding the minimum number of cars) is translated into min-path partitioning of graphs (a path in our graph represents the chain of trips being serviced by the same car). This problem is NP-hard in general graphs (proof by reduction from Hamiltonian Cycle (HC): HC exists if and only if the graph can be covered with 1 path, hence if we find the min number of paths to be 1, the graph has an HC).

\item Since an edge only exists if the trip corresponding to the destination node happens after the source node's, cycles are not possible and a partial ordering is maintained (i.e. the graph is a DAG - directed acyclic graph). Fortunately, there exists a polynomial time algorithm that can solve the path partitioning problem for DAGs.
\end{itemize}

The algorithm works by converting the DAG G into a bipartite graph B where matching in B translates into path partitioning in G:
\begin{itemize}
\item Given G, construct B by creating nodes $i$ in the left part and $i'$ in the right part of B for every node $i$ in G. An \textit{undirected} edge exists between $i$ and $j'$ in B if the directed edge (i,j) exists in G.

\item Running a maximal maximum matching (max-cardinality max-weight) on B would yield the minimum number of cars and their chain of the trip. This can be achieved by modifying weights on the edges by adding N times the maximum score to all edges (so that no augmenting path exists that has fewer edges and higher weight) and then running a maximum matching algorithm (e.g. Hungarian algorithm or Edmond-Karp max-flow equivalent problem). Assume $T = max(score)$ this means every original edge has a value of $[0,T]$, so a matching of size K has a maximum total of $(K * N +1 )* T$. If we take off a match of size $K$ total cost is reduced by a maximum of $(K * N +1 )* T = K N T + T$ and adding a new matching of bigger size K+1 even at the minimum possible weight would be $(K+1) * N *  T = K N T + NT$.  Since the minimum with $K+1$ has definitely more total weight than maximum with $K$ the algorithm would then always choose the longest path.

\item Each chain of trips (path in DAG) has exactly one start and one end (they can be the same), hence the number of chains would be the number of either start or end nodes. Any right part node of B not in the matching is the start of a chain of trips so the number of such nodes is the required number of cars (N - matching\_cardinality). To retrieve the chains for each $i'$ in the right part that is not in the matching, find $i$ in the left part and follow its match chain recursively (i.e. start with $j'$ that $i$ is matched to).
\end{itemize}

\begin{figure}[]
	\centering
    \setlength{\belowcaptionskip}{-1pt}
    \subcaptionbox{} {\includegraphics[width=0.49\columnwidth]{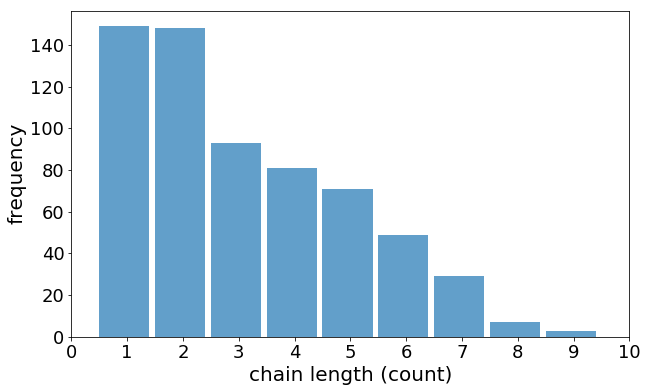}}
    \subcaptionbox{} {\includegraphics[width=0.49\columnwidth]{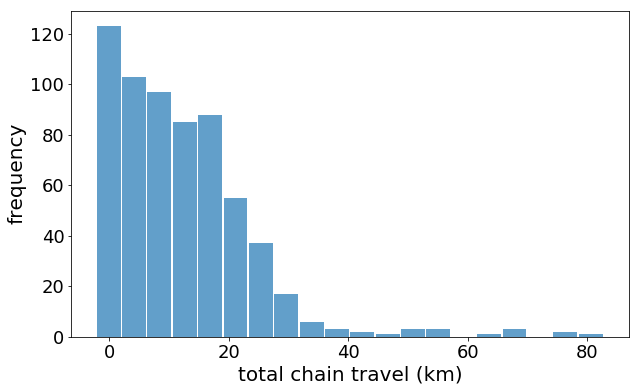}}
    \subcaptionbox{} {\includegraphics[width=0.49\columnwidth]{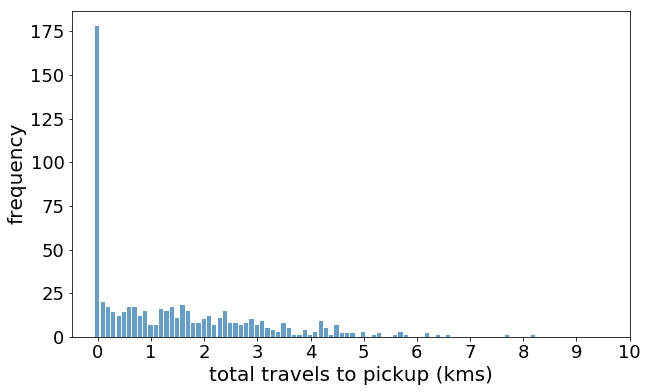}}
    \subcaptionbox{} {\includegraphics[width=0.49\columnwidth]{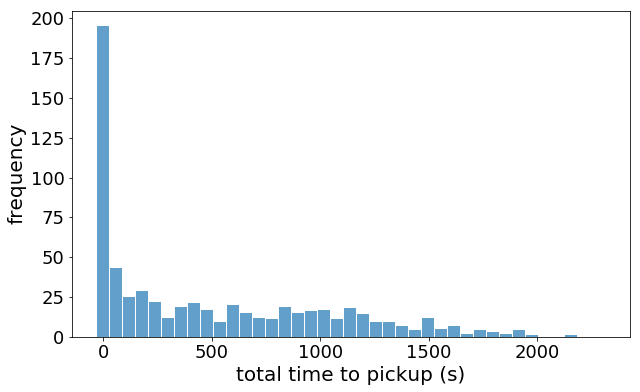}}
    \caption{Carsharing results (a) chain length (b) total km traveled (c) total km traveled to pickup next rider (d) total time spent to pickup next rider.}
    \label{fig:CSH_results}
    \vspace{-6pt}
\end{figure}

The graph constructed by thresholds of 900 seconds and 1.8 KM, results in 38730 edges (2K nodes). The matching cardinality is 1370 which means 630 cars are needed to service 2000 trips with 149 of them not belonging to any chains (singleton trips), leaving 481 chains with the average length of $\approx$3.88. The distribution of chain lengths, chain travels, consecutive pick-up travels and times are presented in Figure \ref{fig:CSH_results} (CDFs are omitted for brevity). Majority of non-singleton trips have chains of size $\leq$4. Approximately 90\% of chains travel <30 KMs, and their total consecutive pick-up distance and time are at most $\approx3.5$ KMs and 1300 seconds. 

\section{Concluding Remarks}\label{sec:conclusion}
In this paper, driven by data, we explored various spatio-temporal characteristics of the trips (of the city of Cologne).
The systematic analysis can be applied to other datasets for understanding such characteristics and for comparison purposes.
Next, We have proposed a similarity score, \textbf{W}eighted \textbf{G}eometric \textbf{M}ean (\textbf{WGM}), that reflects both spatial and temporal similarities with knobs that can be used to favor the spatial or temporal similarity (through weight adjustments).
It is also efficient to calculate.
This score is used in clustering of trips, successfully identifying clusters separated by distance or time.
Next, scenarios of shared transportation including Catch-a-Ride and CarPool are discussed and WGM is used as the criteria for dynamic matching of trips for this purpose.
A comparison of the performance of various similarity scores/distances on the Catch-a-Ride scenario is also given.
WGM is able to capture similar trips in both distance and time dimensions. The weights (of the geometric mean) can be used to capture more similarity in one dimension or the other which gives our score more flexibility while it remains computationally simple compared to other metrics.
Lastly, we formulated the problem of car sharing (autonomous fleet of vehicles) using graph theory and algorithms (with our score as the weights for the graph) which achieves the optimal number of cars required with the minimum spatio-temporal difference between consecutive pick-ups.

To follow up, we plan to confirm the reproducibility of the results on different datasets. A potential dataset is the simulator-generated traces of the trips based on traffic cameras around the globe \cite{enroute,Thakur2012hotplanet}.
Additionally, we plan to find embedding spaces that reflect the similarity of trips. In general, trips are represented through thousands (and not necessarily the same number) of waypoints. This makes detailed comparisons computationally intensive.
Encompassing trips in an embedding space (with a fixed number of dimensions) can ease the problem. Another potential application is in a real-time matching of trips where the details of paths are not shared (e.g. for privacy concerns, or compression). The notion of similarity and its variations can be used for the design of systems with pollution reduction and public safety in mind potentially through participatory sensing and crowdsourcing.
These steps would lead to the concept of profiles of vehicular mobility as a way to describe human mobility behavior.

\begin{acks}
This project was partially supported by NSF grant 1320694 and UF Informatics Institute.
\end{acks}